\documentclass{article}
\usepackage{float}
\usepackage{epsfig}
\usepackage{xcolor}
\usepackage{soul}

\begin{document}

\title{\bf Thin accretion disks around rotating black holes in $4D$ Einstein-Gauss-Bonnet gravity}
\author{{ Mohaddese Heydari-Fard$^{1}$\thanks{Electronic address: m\_heydarifard@sbu.ac.ir}, Malihe Heydari-Fard$^{2}$ \thanks{Electronic address: heydarifard@qom.ac.ir} and Hamid Reza Sepangi$^{1}$\thanks{Electronic address: hr-sepangi@sbu.ac.ir}}\\ {\small \emph{$^{1}$ Department of Physics, Shahid Beheshti University, Evin, Tehran, Iran}}
\\{\small \emph{$^{2}$ Department of Physics, The University of Qom, 3716146611, Qom, Iran}}}

\maketitle

\begin{abstract}
Recently, Kumar and Ghosh have derived Kerr-like rotating black hole solutions in the framework of four-dimensional Einstein-Gauss-Bonnet theory of gravity and investigated the black hole shadow. Using the steady-state Novikov-Thorne model, we study thin accretion disk processes for such rotating black holes including the energy flux, temperature distribution, emission spectrum, energy conversion efficiency as well as the radius of the innermost stable circular orbit. We also study the effects of the Gauss-Bonnet coupling parameter $\alpha$ on these quantities. The results are compared to slowly rotating relativistic Kerr black holes which show that for a positive Gauss-Bonnet coupling, thin accretion disks around rotating black holes in four-dimensional Einstein-Gauss-Bonnet gravity are hotter and more efficient than that for Kerr black holes with the same rotation parameter $a$, while for a negative coupling they are cooler and less efficient. Thus the accretion disk processes may be considered as tools for testing Einstein-Gauss-Bonnet gravity using astrophysical observations.
\vspace{5mm}
\vspace{1mm}
\end{abstract}

\section{Introduction}
Astrophysical objects are expected to grow in mass through accretion. The presence of interstellar matter usually leads to formation of accretion disks around compact objects. An accretion disk is a flattened structure formed by rotating gas which slowly spirals into a massive central body. The gas particles release gravitational energy in the form of heat as they fall into the gravitational potential of the compact object. A fraction of the heat is converted to radiation which is emitted from the inner part of the accretion disk, causing it to cool down. When the emitted radiation reaches radio, optical or X-ray telescopes, it provides the possibility of analyzing its electromagnetic spectrum. The properties of this radiation depend on the geodesic motion of the gas particles which may also be associated with the  structure and nature of the central mass. Therefore important astrophysical information can be obtained from the study of emission spectra of accretion disks.

The standard model of geometrically thin accretion disks, first proposed by Shakura and Sunyaev in 1973, is based on Newtonian approach \cite {Shakura} and was extended to the case of general relativity (GR) later on by Novikov and Thorne \cite{Novikov}. In this model the mass accretion rate is assumed to be constant and independent of the radius of the disk, that is, the disk is in a steady state. Also, it is assumed that the accreting matter has Keplerian motion which requires the central mass to be devoid of a strong magnetic field. Moreover, the radiation emitted from the disk is considered as black body radiation, resulting from thermodynamic equilibrium of the disk. The properties of the energy flux over the disk surface was  analyzed in \cite{Page} and \cite{Thorne}. In this analysis the radiative efficiency, in the sense of the capability of the central compact object to convert rest mass into outgoing radiation via the accretion process, was also computed. Thin accretion disk properties in modified theories of gravity such as $f(R)$ gravity \cite{FR1}--\cite{FR3}, scalar-tensor-vector gravity \cite{SVT}, Einstein-Maxwell-dilaton theory \cite{EMd1}--\cite{EMd2}, Einstein-scalar-Gauss-Bonnet gravity\cite{EdGB1}--\cite{EdGB2}, Chern-Simons \cite{Chern} and Horava-Lifshitz \cite{Horava} gravity have been studied in the past. In higher-dimensional gravity models such as Kaluza-Klein  and brane-world modeles, thin accretion disks have been investigated in \cite{Kaluza}--\cite{brane2}. Also the study of thin accretion disks based on the Novikov-Thorne model, in the space-times of wormholes, neutron, boson and fermion stars and naked singularities have been carried out in \cite{WH1}--\cite{nk3}, respectively. For study of thin accretion disks in $4D$ Einstein-Gauss-Bonnet (EGB) gravity, see \cite{EGB}. In this paper we propose to extend the latter to the case of rotating black hole (BH) solutions of  $4D$ EGB gravity, since astrophysical BHs are expected to be rapidly rotating due to the accretion effects.

In recent years, gravitational theories with higher-order curvature corrections to the Einstein-Hilbert action of GR have been the focus of attention  since such curvature corrections appear in quantum gravity and string theory. In higher-dimensional space-times, $D>4$, the low energy limit of heterotic string theory predicts a second-order curvature correction to Einstein-Hilbert action which is the well known GB term. This term is a specific combination of higher-order curvature invariants which is a natural extension of Einstein's GR in a $D$-dimensional space-time with $D-4$ extra dimensions. In $D=4$ the GB term is a topological invariant and does not contribute to the gravitational field equations. This is no longer the case when a scalar field is coupled to the GB term through a regular coupling function, a well-known example of which is the Einstein-dilaton-Gauss-Bonnet gravity \cite{Kanti}.

However, recently a $4D$ EGB gravity has been proposed by Glavan and Lin \cite{Glavan} where by re-scaling the GB coupling constant $\alpha$ according to $\alpha\rightarrow\frac{\alpha}{D-4}$ and taking the limit $D\rightarrow 4$, the GB term does contribute to the field equations and thus circumvents the Lovelock theorem. This theory preserves the number of degrees of freedom and avoids the Ostrogradsky instability. Also, they have constructed a static and spherically symmetric BH solution which is free from the singularity problem. Note that such BH solution has been obtained earlier in a semi-classical gravity framework with conformal anomaly \cite{Cai}, but this $4D$ EGB gravity is a classical modified theory of gravity in equal footing with GR. However, several criticisms on the regularization process used in \cite{Glavan} have come into fore \cite{c1}--\cite{c7}. It is argued that taking the limit $D\rightarrow 4$ may not be consistent and the theory is not well-defined in four-dimensions. At the same time, some prescriptions including compactification of $D$-dimensional EGB gravity \cite{n1}--\cite{n2}, introducing a counter term into the action \cite{n3}--\cite{n4} and breaking the temporal diffeomorphism invariance \cite{n5} have been suggested as remedies to address this problem and to obtain a consistent EGB gravity. It is important to note that in these consistent theories the spherically symmetric BH solutions obtained in \cite{Glavan} are still valid and worthy of study. For instance, charged and rotating $4D$ EGB BH solutions \cite{charge}--\cite{rotating2}, BH solutions in Lovelock gravity \cite{love1}--\cite{love2}, BH solutions surrounded by clouds of strings \cite{string}, Bardeen BHs \cite{bardeen}, Hayward BHs \cite{hayward}, spherically symmetric and thin shell wormhole solutions \cite{wormhole1}--\cite{wormhole2} and relativistic stars \cite{star} have been extensively studied. Also, a large number of interesting aspects of the theory including geodesic motion and shadow \cite{isco}--\cite{shadow2}, strong and weak gravitational lensing \cite{lensing1}--\cite{lensing4}, quasinormal modes of BHs \cite{QNM1}--\cite{QNM5}, instability of (A)dS BHs \cite{stability1}--\cite{stability3}, thermodynamics and phase transition \cite{th1}--\cite{th4}, Hawking radiation \cite{Hawking1}--\cite{Hawking2} and new quark stars \cite{star1}--\cite{star2} have also been studied. For further references on $4D$ EGB gravity see \cite{a1}--\cite{a15}.

The structure of the paper is as follows. In section 2, we review the geodesic motion of test particles moving in a general stationary axisymmetric space-time. In section 3 we present the Novikov-Thorne model as the standard framework for studying  geometrically thin accretion disks. The novel $4D$ EGB gravity is introduced and the electromagnetic properties of thin accretion disks around rotating EGB BHs is studied in section 4. Finally, we present the conclusions in section 5.

 \section{Generic rotating space-times and geodesic equations}
The line element of a generic stationary and axisymmetric space-time is given by
\begin{equation}
ds^2=g_{tt}dt^2+2g_{t\phi}dtd\phi+g_{rr}dr^2+g_{\theta\theta}d\theta^2+g_{\phi\phi}d\phi^2,
\label{1}
\end{equation}
where we assume that the metric coefficients $g_{tt}, g_{rr}, g_{\theta\theta}$, $g_{\phi\phi}$ and $g_{t\phi}$ are functions of $r$ and $\theta$ coordinates. Since the above metric is independent of $t$ and $\phi$ coordinates, we have two constants of motion, namely the energy and the angular momentum per unit rest-mass, $\tilde{E}$ and $\tilde{L}$, as follows
\begin{equation}
g_{tt}\dot{t}+g_{t\phi}\dot{\phi}=-\tilde{E},
\label{2}
\end{equation}
\begin{equation}
g_{t\phi}\dot{t}+g_{\phi\phi}\dot{\phi}=\tilde{L},
\label{3}
\end{equation}
where a dot denotes derivative with respect to the affine parameter $\tau$. Using equations (\ref{2}) and (\ref{3}) we find $t$ and $\phi$ components of the $4$-velocity $\dot{x}^{\mu}$ as
\begin{equation}
\dot{t}=\frac{\tilde{E}g_{\phi\phi}+\tilde{L}g_{t\phi}}{g_{t\phi}^2-g_{tt}g_{\phi\phi}},
\label{4}
\end{equation}
\begin{equation}
\dot{\phi}=-\frac{\tilde{E}g_{t\phi}+\tilde{L}g_{tt}}{g_{t\phi}^2-g_{tt}g_{\phi\phi}}.
\label{5}
\end{equation}
From the normalization condition, $g_{\mu\nu}\dot{x}^{\mu}\dot{x}^{\nu}=-1$, we obtain
\begin{equation}
g_{rr} \dot{r}^2+g_{\theta\theta}\dot{\theta}^2=V_{\rm eff}(r,\theta),
\label{6}
\end{equation}
where the effective potential reads
\begin{equation}
V_{\rm eff}(r,\theta)=-1+\frac{\tilde{E}^2g_{\phi\phi}+2\tilde{E}\tilde{L}g_{t\phi}+\tilde{L}^2g_{tt}}{g_{t\phi}^2-g_{tt}g_{\phi\phi}}.
\label{7}
\end{equation}
For circular orbits in the equatorial plane $(\theta=\pi/2)$ with $\dot{r}=\dot{\theta}=0$ we have $V_{\rm eff}(r)=0$, and $\ddot{r}=\ddot{\theta}=0$ which require $V_{\rm eff,r}=0$ and $V_{\rm eff,\theta}=0$, respectively. Using metric (\ref{1}) and these conditions we can find the specific energy and specific angular momentum for the test particles in circular orbits. However, a more efficient way is to use geodesic equations. The radial component of the geodesic equation with conditions $\dot{r}=\dot{\theta}=\ddot{r}=0$ for equatorial circular orbits leads to the angular velocity $\Omega=\dot{t}/\dot{\phi}$ as follows
\begin{equation}
\Omega_{\pm}=\frac{-g_{t\phi,r}\pm\sqrt{(g_{t\phi,r})^2-g_{tt,r}g_{\phi\phi,r}}}{g_{\phi\phi,r}},
\label{8}
\end{equation}
where the upper sign denotes co-rotating orbits with angular momentum parallel to the BH spin, while the lower sign refers to counter-rotating orbits with angular momentum antiparallel to the spin of the BH. Then, from $g_{\mu\nu}\dot{x}^{\mu}\dot{x}^{\nu}=-1$ with $\dot{r}=\dot{\theta}=0$ and equations (\ref{2}) and (\ref{3}), the specific angular momentum ${\tilde{L}}$ and the specific energy ${\tilde{E}}$, for a particle on a circular orbit in the gravitational potential of a massive object can be written as
\begin{equation}
{\tilde{E}}=-\frac{g_{tt}+g_{t\phi}\Omega}{\sqrt{-g_{tt}-2g_{t\phi}\Omega-g_{\phi\phi}\Omega^2}},
\label{9}
\end{equation}
\begin{equation}
{\tilde{L}}=\frac{g_{t\phi}+g_{\phi\phi}\Omega}{\sqrt{-g_{tt}-2g_{t\phi}\Omega-g_{\phi\phi}\Omega^2}}.
\label{10}
\end{equation}
For test particles in the gravitational potential of a central body, the innermost stable circular orbit known as the ISCO radius is defined as
\begin{equation}
V_{\rm eff,rr}\mid_{r=r_{\rm isco}}=\frac{1}{g_{t\phi}^2-g_{tt}g_{t\phi}}\left[{\tilde{E}^2g_{\phi\phi,rr}}
+2\tilde{E}\tilde{L}g_{t\phi,rr}+{\tilde{L}^2g_{tt,rr}}-\left(g_{t\phi}^2-g_{tt}g_{\phi\phi}\right)_{,rr}\right]\mid_{r=r_{\rm isco}}=0.
\label{11}
\end{equation}
Since the equatorial circular orbits are unstable for $r<r_{\rm isco}$,  $r_{\rm isco}$ determines the inner edge of thin accretion disks in the Novikov-Thorne model.

\section{Thin accretion disks around compact objects}
Let us now review the physical properties of thin accretion disks that we will need in our calculations, such as energy flux emitted by the disk,  $F(r)$, temperature distribution, $T(r)$, Luminosity spectra, $L(\nu)$ and  efficiency $\epsilon$. The standard framework in the explanation of thin accretion disk processes is the Novikov-Thorne \cite{Novikov} model which is a generalization of that of the Shakura-Sunyaev  \cite{Shakura}. There are various versions of the model, but we start by stating some typical assumptions as follows:
\begin{enumerate}
  \item The space-time describing the central massive object is stationary, axisymmetric and asymptotically flat.
  \item The self-gravity of the disk is negligible so that disk's mass has no effect on the background metric.
  \item The accretion disk is geometrically thin, namely its vertical size $h$, is negligible compared to its horizontal size, $h\ll r$.
  \item The ISCO radius determines the inner edge of the disk and orbiting particles around the compact central object move between $r_{\rm isco}$ and the outer edge $r_{\rm out}$.
  \item The disk surface is perpendicular to the BH spin, namely the accretion disk lies in the equatorial plane of the accreting compact object.
  \item The emitted electromagnetic radiation from the disk surface is assumed to have a black body spectrum resulting from  hydrodynamic and thermodynamic equilibrium of the disk.
  \item The disk is in a steady-state, namely the mass accretion rate, $\dot{M}_0$, does not change with time.
\end{enumerate}
The radiant energy flux over the disk surface can be obtained from the conservation equations of rest mass, energy, and the angular momentum of the disk particles according to \cite{Novikov}--\cite{Page}
\begin{equation}
F(r)=-\frac{\dot{M}_{0}\Omega_{,r}}{4\pi \sqrt{-g}\left(\tilde{E}-\Omega \tilde{L}\right)^2}\int^r_{r_{\rm isco}}\left(\tilde{E}-\Omega \tilde{L}\right) \tilde{L}_{,r}dr,
\label{12}
\end{equation}
where $\dot{M}_{0}$ is the mass accretion rate. Due to the thermal equilibrium of the disk, as was mentioned above, we can use the Stefan-Boltzmann law to find the disk temperature
\begin{equation}
F(r)=\sigma_{\rm SB} T(r)^4,\label{13}
\end{equation}
where $\sigma_{\rm SB}=5.67\times10^{-5}\rm erg$ $\rm s^{-1} cm^{-2} K^{-4}$ is the Stefan-Boltzmann constant. Also, the observed luminosity $L(\nu)$ of a thin accretion disk has a red-shifted black body spectrum given by  \cite{Torres}
\begin{equation}
L(\nu)=4\pi d^2 I(\nu)=\frac{8\pi h \cos\gamma}{c^2}\int_{r_{\rm in}}^{r_{\rm out}}\int_0^{2\pi}\frac{ \nu_e^3 r dr d\phi}{\exp{[\frac{h\nu_e}{k_{\rm B} T}]}-1
},\label{14}
\end{equation}
where $d$ is the distance to the disk center, $\gamma$ is the disk inclination angle (which we will set to be zero), and $r_{\rm in}$ and $r_{\rm out}$ are inner and outer radii of the edge of the disk, respectively. The Planck and Boltzmann constants are respectively presented by $h$ and $k_{\rm B}$, and $\nu_{e}=\nu(1+z)$ is the emitted frequency where the redshift factor $z$ can be written as
\begin{equation}
1+z=\frac{1+\Omega r\sin\phi\sin\gamma}{\sqrt{-g_{tt}-2g_{t\phi}\Omega-g_{\phi\phi}\Omega^2}}.\label{15}
\end{equation}
Another important quantity is the radiative efficiency which indicates the capability of the BH to convert rest mass into radiation. When  absorption by the BH is negligible, the Novikov-Thorne efficiency is given by \cite{Thorne}
\begin{equation}
\epsilon=1-\tilde{E}_{\rm isco},\label{16}
\end{equation}
where $\tilde{E}_{\rm isco}$ is the specific energy of test particles measured at the ISCO radius.

\section{Accretion disk processes in $4D$ EGB gravity}
\subsection{Rotating $4D$ EGB BHs}

The action of EGB gravity in $D$-dimension is defined as follows
\begin{equation}
{\cal S}_{\rm EGB}=\int d^{D}x \sqrt{-g}({\cal L}_{\rm EH}+\alpha {\cal L}_{\rm GB}),
\label{17}
\end{equation}
with
\begin{eqnarray*}
{\cal L}_{\rm EH}= R, \quad {\cal L}_{\rm GB}=R^{\mu\nu\rho\sigma}R_{\mu\nu\rho\sigma}-4R^{\mu\nu}R_{\mu\nu}+R^2,
\end{eqnarray*}
where $g$, and $\alpha$ are the determinant of $g_{\mu\nu}$ and the GB coupling constant, respectively with $R$, $R_{\mu\nu}$ and $R_{\mu\nu\rho\sigma}$ being the Ricci scalar,
Ricci tensor and Riemann tensor of the space-time. As we mentioned before, in $D=4$ the GB term is a total derivative and so does not contribute to the Einstein field equations. However, recently Glavan and Lin proposed a novel $4D$ EGB gravity by rescaling the GB coupling $\alpha$ as $\frac{\alpha}{D-4}$ and taking the limit $D\rightarrow4$ and derived the static spherically symmetric BH solutions of the theory \cite{Glavan}.  Varying the action (\ref{17}) with respect to the metric tensor $g_{\mu\nu}$ gives the gravitational field equations
\begin{equation}
G_{\mu\nu}+\alpha H_{\mu\nu}=0,
\label{field1}
\end{equation}
where $G_{\mu\nu}$ is the Einstein tensor and $H_{\mu\nu}$ is given by
\begin{equation}
H_{\mu\nu}=2\left(RR_{\mu\nu}-2R_{\mu\sigma}R^{\sigma}_{\nu}-2R_{\mu\sigma\nu\rho}R^{\sigma\rho}-
R_{\mu\sigma\rho\delta}R^{\sigma\rho\delta}_{\,\,\,\,\,\,\,\,\,\,\nu}\right)-\frac{1}{2}{\cal L}_{\rm GB}g_{\mu\nu},
\label{field2}
\end{equation}

Here, it is worth mentioning that the construction of rotating $4D$ EGB BH solutions by solving the vacuum field equations is a formidable task. There exists no exact solution of any equation for EGB rotating BHs. All that has been done is some insightful guesswork to write the metric which has all the desired properties of a rotating BH. Indeed, recently Kumar and Ghosh, applying the Newman-Janis algorithm to a non-rotating BH in $4D$ EGB gravity, have constructed the metric for a stationary and axially symmetric rotating BH \cite{rotating1}, which in the Boyer-Lindqist coordinates is given by
\begin{eqnarray}
ds^2&=&-\left(\frac{\Delta-a^2\sin^2\theta}{\Sigma}\right)dt^2+\frac{\Sigma}{\Delta}{dr^2}-2a\sin^2\theta\left(1-\frac{\Delta-a^2\sin^2\theta}{\Sigma}\right)dt d\phi+\Sigma d\theta^2\nonumber\\
&+&\sin^2\theta\left[\Sigma+a^2\sin^2\theta\left(2-\frac{\Delta-a^2\sin^2\theta}{\Sigma}\right)\right] d\phi^2,\label{18}
\end{eqnarray}
with
\begin{equation}
\Delta=r^2+a^2+\frac{r^4}{2\alpha}\left[1-\sqrt{1+\frac{8\alpha M}{r^3}}\right], \quad \Sigma=r^2+a^2\cos^2\theta,\label{19}
\end{equation}
where $M$ is the BH mass and $a$ is the rotation parameter. The above metric, in the limit $\alpha\rightarrow0$ or large $r$, becomes the rotating solution of GR which is described by the Kerr metric. In the limit $a=0$ it reduces to the static spherically symmetric solution of $4D$ EGB gravity and also for both $\alpha\rightarrow0$ and $a=0$ it is exactly the Schwarzschild metric.

Using the existence condition of the horizon, one can obtain a bound on the GB coupling $\alpha$ for a specific value of the rotation parameter $a$. In figure 1, we have plotted $\Delta(r)$ as a function of the radial coordinate $r$ for different values of $a$ with positive $\alpha$. As is clear, in each panel, for some values of the GB coupling $\alpha$, there are two distinct horizons, i.e. the inner Cauchy horizon, $r_{-}$, and the outer event horizon, $r_{+}$, so that $r_{-}<r_{+}$. For a critical value of $\alpha=\alpha_{\rm c}$ the two horizons coincide $r_{-}=r_{+}$, and we get an extremal BH \footnote{Note that we have set $16\pi G=1$, so the critical values of $\alpha_{\rm c}$ for the extremal BHs are different from those in \cite{rotating1}.}. Also, for $\alpha >\alpha_{\rm c}$ there are no BH space-time. The figure shows the effect of the GB coupling parameter $\alpha$ on the event horizon. For a fixed value of $a$, decreasing the GB coupling $\alpha$ causes the event horizon radius to increase so that for the Kerr BH ($\alpha\rightarrow0$) in GR it is larger than that of rotating EGB BHs. Note that although the GB coupling is identified as the inverse string tension and should be positive, it was found that \cite{isco}  for $-8\leq\alpha<0$ there always exist a BH solution and the singular behavior of the solution is hidden inside the horizon. Moreover, in contrast to the case of positive $\alpha$, equation $\Delta(r)=0$ has only one positive real root and thus there only exists one BH horizon for negative GB coupling. In the following, we will consider the BH solutions with $-8\leq\alpha<1$.

\begin{figure}[H]
\centering
\includegraphics[width=3.0in]{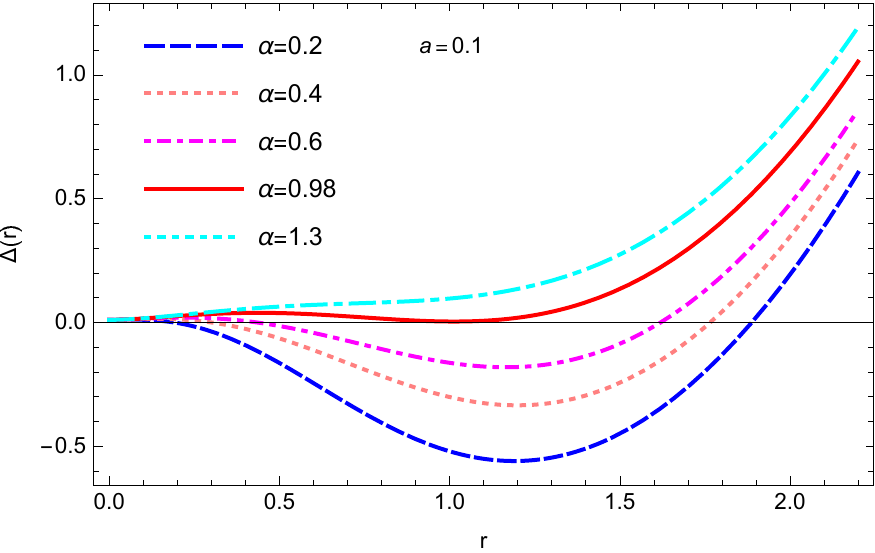}
\includegraphics[width=3.0in]{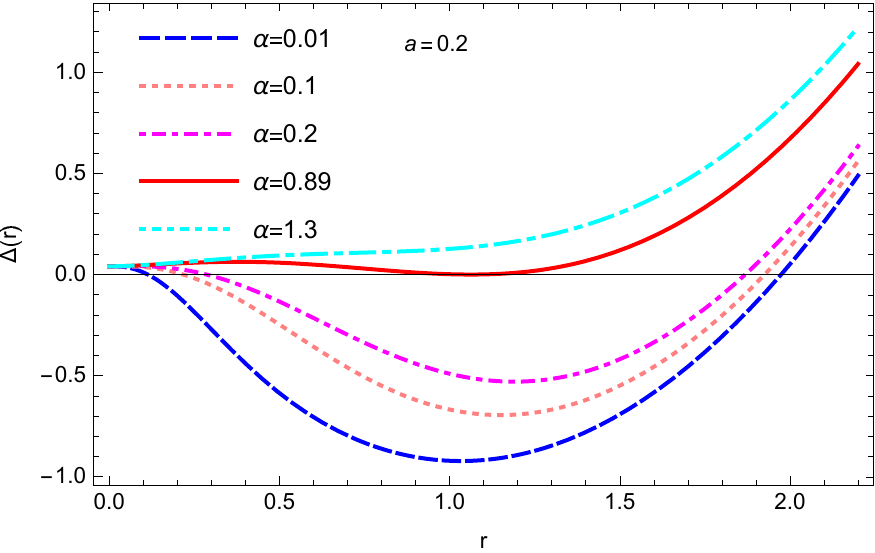}\\
\includegraphics[width=3.0in]{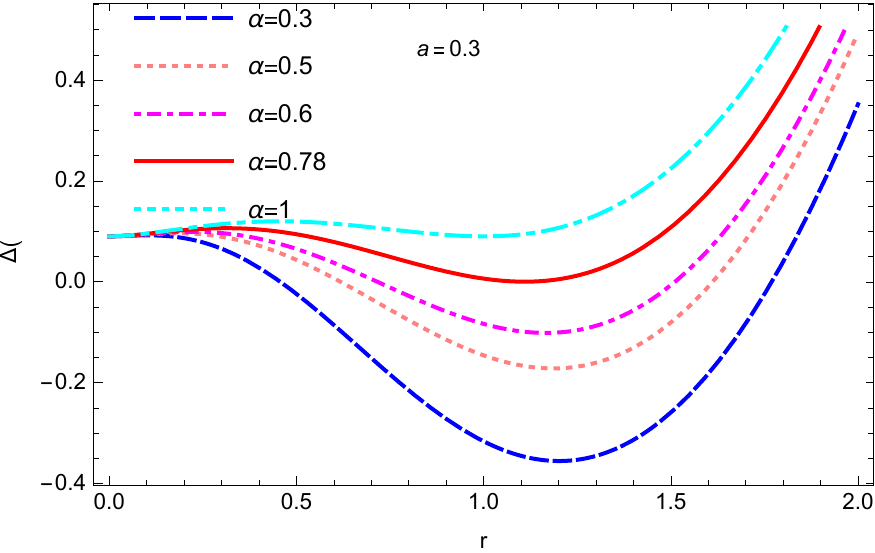}
\includegraphics[width=3.0in]{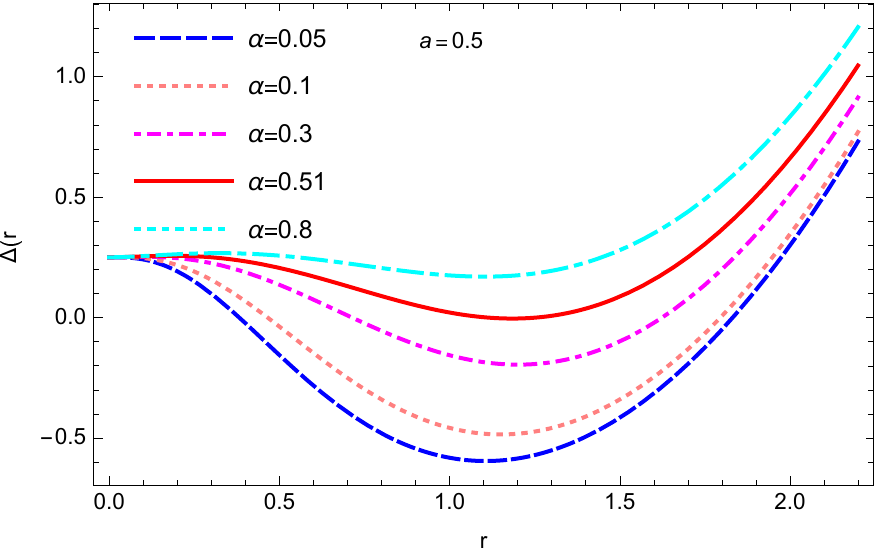}
\caption{\footnotesize The behavior of  horizons as a function of the radial coordinate $r$ for different values of the GB coupling parameter $\alpha$ with a=0.1 (top-left panel), a=0.2 (top-right panel), a=0.3 (bottom-left panel) and a=0.5 (bottom-right panel), respectively. In each panel the solid curve corresponds to the extremal BH.}
\label{horizon}
\end{figure}

\subsection{Thin accretion disk properties around rotating EGB BHs}
Now, we aim to investigate thin accretion disk properties around rotating EGB BHs. Using equations (\ref{8})-(\ref{10}) we obtain the specific energy, specific angular momentum and angular velocity  as follows
\begin{equation}
\tilde{E}=\frac{1+\frac{h_1}{r^2}-\frac{2ah_1}{r^2}\Omega}{\sqrt{1+\frac{h_1}{r^2}-\frac{4ah_1}{r^2}\Omega-\Omega^2[r^2+a^2(1-\frac{h_1}{r^2})]}},\label{20}
\end{equation}
\begin{equation}
\tilde{L}=\frac{\frac{2ah_1}{r^2}+[r^2+a^2(1-\frac{h_1}{r^2})]\Omega}{\sqrt{1+\frac{h_1}{r^2}-\frac{4ah_1}{r^2}\Omega-\Omega^2[r^2+a^2(1-\frac{h_1}{r^2})]}},\label{21}
\end{equation}
where
\begin{equation}
\Omega=\frac{-a(h_2+\frac{4h_1}{r^3})+\sqrt{a^2(h_2+\frac{4h_1}{r^3})^2-\frac{h_2h_3}{(12M\alpha)^2}(24M\alpha r-a^2h_2h_3)}}{2r-\frac{a^2}{2}(h_2+\frac{4h_1}{r^3})},\label{22}
\end{equation}
with
$$
h_1=\frac{r^4}{2\alpha}\left[1-\sqrt{1+\frac{8\alpha M}{r^3}}\right],
$$
$$
h_2=\frac{12M}{r^2\sqrt{1+\frac{8M\alpha}{r^3}}},
$$
$$
h_3=2M\alpha+\frac{2\alpha}{r}h_1.
$$
The ISCO equation (\ref{11}) is also given by
\begin{equation}
r^2-6Mr+8aM^{1/2}r^{1/2}-3a^2+16M\alpha r^{-1}+96\alpha aM^{3/2}r^{-5/2}+72\alpha aM^{5/2}r^{-7/2}=0,\label{isco1}
\end{equation}
where we have only kept linear terms in GB coupling. However, even if we were only to keep the term $16M\alpha r^{-1}$ containing the lowest-order of  $M$ and $\alpha$ and ignore the last two terms, the resulting equation would have no analytic solution and had to be solved numerically to obtain the ISCO radius. It is also seen that in the case of $a=\alpha=0$ it reduces to the Schwarzschild BH with $r_{isco}=6M$ and in the case of $\alpha=0$, it reduces to equation (2.20) of \cite{Bardeen} for the Kerr BHs that has a solution of the form
\begin{equation}
r_{isco}=M\left[3+Z_2\mp\sqrt{(3-Z_1)(3+Z_1+2Z_2)}\right],\label{kerr}
\end{equation}
with
$$
Z_1\equiv1+\left(1-\frac{a^2}{M^2}\right)^{1/3}\left[\left(1+\frac{a}{M}\right)^{1/3}+\left(1-\frac{a}{M}\right)^{1/3}\right],
$$
$$
Z_2\equiv\sqrt{3\frac{a^2}{M^2}+Z_1^2},
$$
which, again, the upper and lower sign refer to co-rotating and counter-rotating orbits, respectively. Moreover, the results for static $4D$ EGB BHs can be recovered in the case of $a=0$.
\begin{figure}[H]
\centering
\includegraphics[width=3.0in]{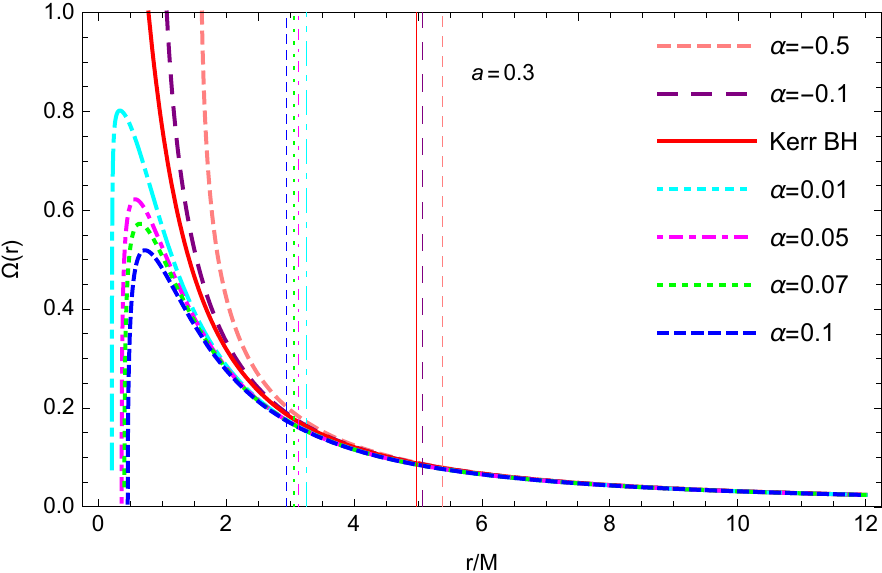}
\includegraphics[width=3.0in]{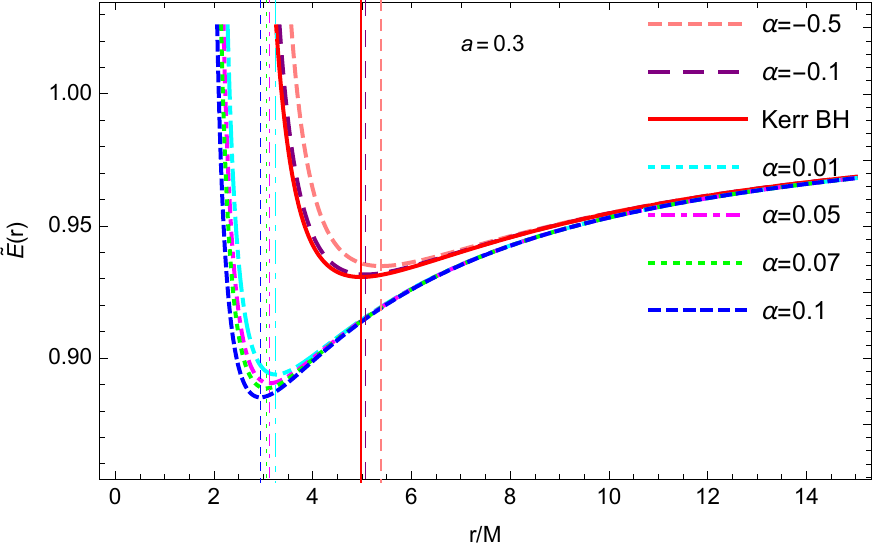}\\
\includegraphics[width=3.0in]{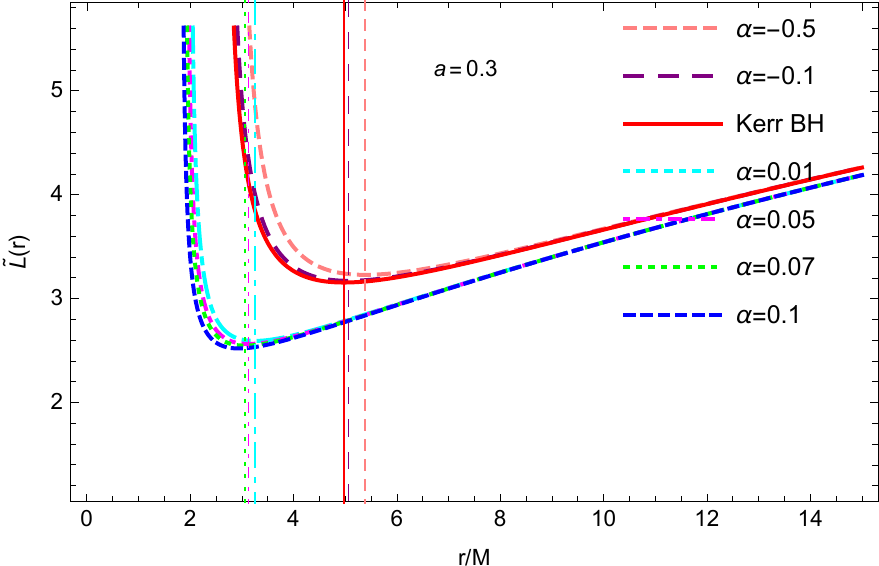}
\caption{\footnotesize The angular velocity $\Omega(r)$ (top-left panel), specific energy $\tilde{E}(r)$ (top-right panel) and  specific angular momentum $\tilde{L}(r)$ (bottom panel) of a rotating $4D$ EGB BH with total mass $M=2.5\times10^6M_{\odot}$ shown as a function of the radial coordinate $r$ for different values of $\alpha$ and compared to a slowly rotating Kerr BH. The rotation parameter is set to $a=0.3$. The vertical lines represent the location of the ISCO in each case.}
\label{potential1}
\end{figure}

Figure 2 shows the specific energy, specific angular momentum and angular velocity for rotating EGB BHs for $a=0.3$ and different values of $\alpha$. It also shows a comparison with the corresponding results for a slowly rotating Kerr BH in GR. We see that a positive GB coupling causes the above quantities to decreases in comparison to Kerr BHs and as the value of $\alpha$ increases, deviation from GR increases too, while a negative $\alpha$ causes them to increase compared to those of a Kerr BH. Also, the behavior of $V_{\rm eff,rr}$ for $a=0.3$ and different values of $\alpha$ is  plotted in figure 3 to present the dependence of $r_{\rm isco}$ on the GB coupling. As is clear, for positive $\alpha$ its zero shifts to smaller radii with increasing GB coupling parameter and thus the ISCO is smaller than the Kerr BH, while for $\alpha<0$ the ISCO radius of rotating EGB BHs is larger than that for the Kerr BH. The figure shows that for $\alpha =0.2$ equation (\ref{11}) has no solution and $V_{\rm eff,rr}$ is negative for any $r$, resulting in particles moving around the BH having stable circular orbits for $r>r_+$.

We have also numerically obtained the ISCO radius from equation (\ref{11}) for different values of rotation parameter $a$. Calculations show that for $a\geq0.4$ the ISCO equation has no solution for any values of $\alpha<\alpha_{\rm c}$ (there is no BH solutions for $\alpha>\alpha_{\rm c}$). Thus in Table 1 we have presented results for $a=0.1, 0.2, 0.3$, corresponding to the slow rotation limit with different values of $\alpha$ to compare the results with a slowly rotating Kerr BH. Note that the slow rotation approximation means that the BH spin is much smaller than the BH mass $a/M\ll1$. For the Kerr space-time the slow rotation limit requires the event horizon of the Kerr BH coincides with the location of the horizon of a Schwarzschild BH, $r_{H}=2M$. Similarly, it is easy to show that the event horizon of rotating $4D$ EGB BHs in the slow rotation limit matches that {of the horizon of static EGB BHs.

Table 1 shows that for a fixed value of the rotation parameter the effect of the GB coupling is to decrease the ISCO radius of rotating EGB BHs compared to the Kerr BH, so that with increasing $\alpha$ the ISCO radius decreases. This is due to the fact that the GB term is a candidate for dark energy and effectively counteracts gravity and thus causes the ISCO radius to lie at a smaller distance from the BH. We also see that in a Kerr space-time, by increasing $a$ for a co-rotating disk, the radiative efficiency increases from $6\%$ to about $7\%$. Similar to a Kerr BH in the case of rotating EGB BHs, at a given value of $\alpha$, increasing $a$ causes the radiative efficiency to increase for co-rotating disks. For instance, in the case of $\alpha=0.07$, increasing the spin parameter results in the efficiency to increase from $6.5\%$ for $a=0.1$ to $11\%$ for $a=0.3$. However, the radiative efficiency is a decreasing function of the spin for a counter-rotating disk.

\begin{figure}[H]
\centering
\includegraphics[width=3.0in]{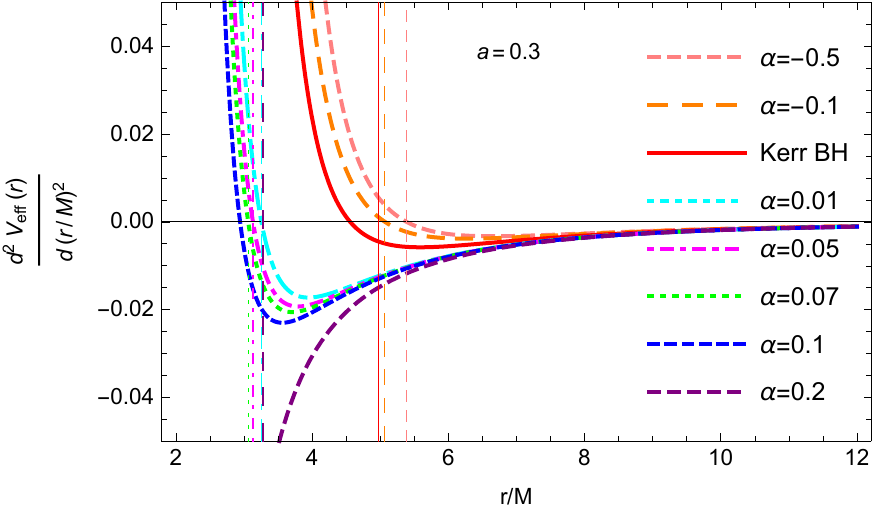}
\caption{\footnotesize The second derivative of the effective potential of rotating $4D$ EGB BHs for $a=0.3$ and different values of the GB coupling parameter $\alpha$. The vertical lines represent the location of the ISCO in each case.}
\label{potential2}
\end{figure}

Now, in order to display the flux distribution for rotating EGB BHs, we consider a BH with a total mass of $M=2.5\times 10^{6}M_{\odot }$ and a mass accretion rate of $\dot{M}=2\times10^{-6}M_{\odot}\rm yr^{-1}$. In units of the Eddington accretion rate we have  $\dot{M}=3.36\times10^{-4}{\dot M_{\rm Edd}}$ which is in the range for supermassive BHs. The Eddington luminosity is defined as the maximum luminosity of an object and can be obtained when there is the equality between the outward radiation pressure  and the gravitational force acting inward. For an object with mass $M$, the Eddington luminosity is given by
\begin{equation}
L_{\rm Edd}=1.26\times10^{38}\left(\frac{M}{M_{\odot}}\right)\frac{erg}{s},
\label{edd}
\end{equation}
and for an accreting BH, the Eddington mass accretion rate is defined as $\dot M_{\rm Edd}c^2\equiv L_{\rm Edd}$ \cite{book}. It is found that for geometrically thin accretion disks with the inner edge located at the ISCO radius, the accretion luminosity should be between 5\% to 30\% of the Eddington limit \cite{ed1}--\cite{ed4}. So, here we have chosen the values of $M$ and $\dot{M}$ that are below the Eddington limit and represent a supermassive BH. For instance, a well known astronomical source is SgrA$^{*}$, a supermassive BH at the center of the Milky Way with a mass of order $M=4.1\times 10^{6}M_{\odot}$ and with an estimated rate $\dot{M}\sim10^{-9}-10^{-7}M_{\odot}\rm yr^{-1}$ \cite{sgr}.

The energy flux over the surface of the disk for $a=0.1$ and $a=0.3$ and different values of $\alpha$ is plotted in figure 4. As can be seen, for positive $\alpha$ the energy emanating from the disk in $4D$ EGB BHs is larger than that for the Kerr BH, while for negative $\alpha$ the energy flux is smaller than that for the Kerr  BH. For a fixed value of the rotation parameter, increasing the GB parameter causes the energy flux to also increase. The effect of the GB coupling becomes more prominent as the BHs are rotating faster so that for larger values of the rotation parameter, the same increase in the GB coupling leads to higher values for the energy flux and shifts the ISCO radius to lower and lower radii. Moreover, we see that for larger values of $\alpha$, the radial position of the maximal flux shifts to lower radii, approaching the ISCO location. This shift of the maximal locations to lower radii clearly shows that most of the radiation is emitted from the inner part of the accretion disk.

The behavior of disk temperature is shown in figure 5 where the same features can be seen. With increasing $\alpha$, the disk temperature increases and the maximum values shift closer to the inner edge of the disk. Moreover,  for positive values of the GB coupling, disks rotating around a $4D$ EGB BHs are much hotter than the disks around Kerr BHs, while for negative values they are cooler and less efficient.

\begin{figure}[H]
\centering
\includegraphics[width=3.0in]{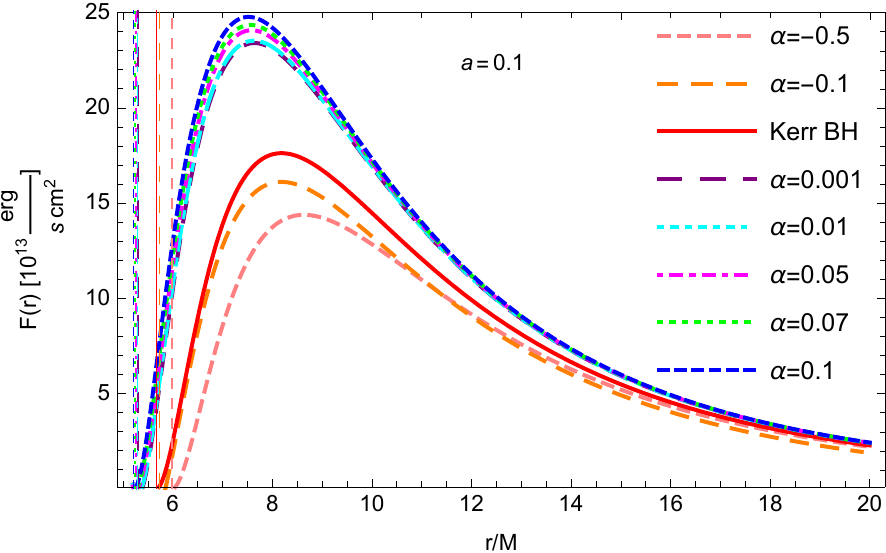}
\includegraphics[width=3.0in]{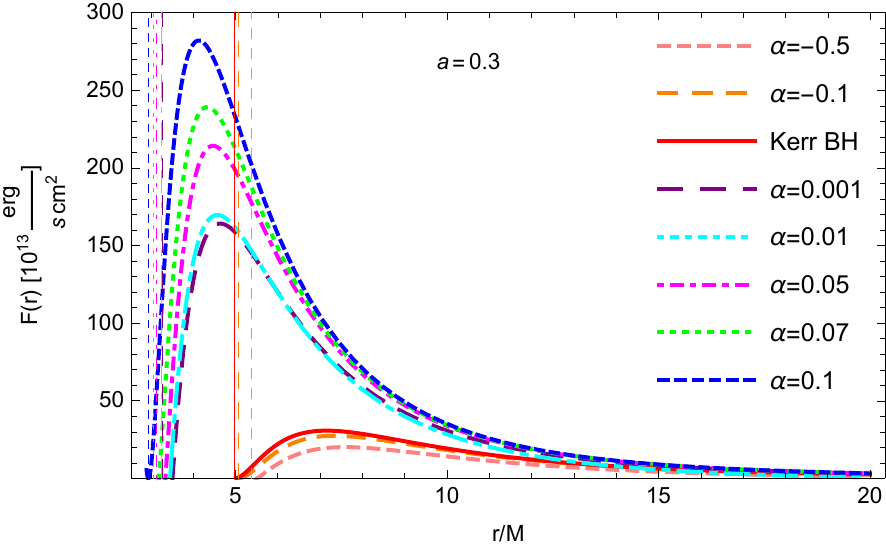}
\caption{\footnotesize The energy flux $F(r)$ from a disk around a rotating BH in EGB gravity with the mass accretion rate $\dot{M}=2\times10^{-6}M_{\odot}\rm yr^{-1}$, for different values of the GB coupling constant $\alpha$. The rotation parameter is set to $a=0.1$ (left panel) and $a=0.3$ (right panel), respectively. In each panel the solid curve corresponds to a slowly rotating Kerr BH. The vertical lines represent the location of the ISCO in each case.}
\label{flux}
\end{figure}

\begin{figure}
\centering
\includegraphics[width=3.0in]{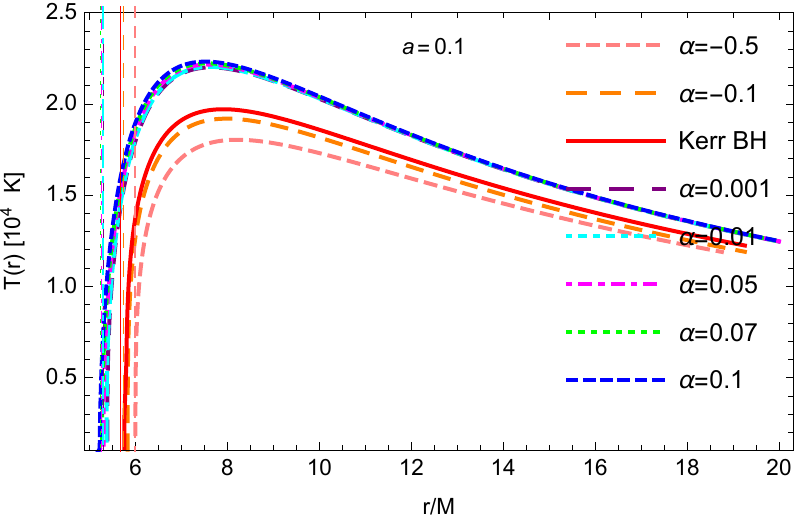}
\includegraphics[width=3.0in]{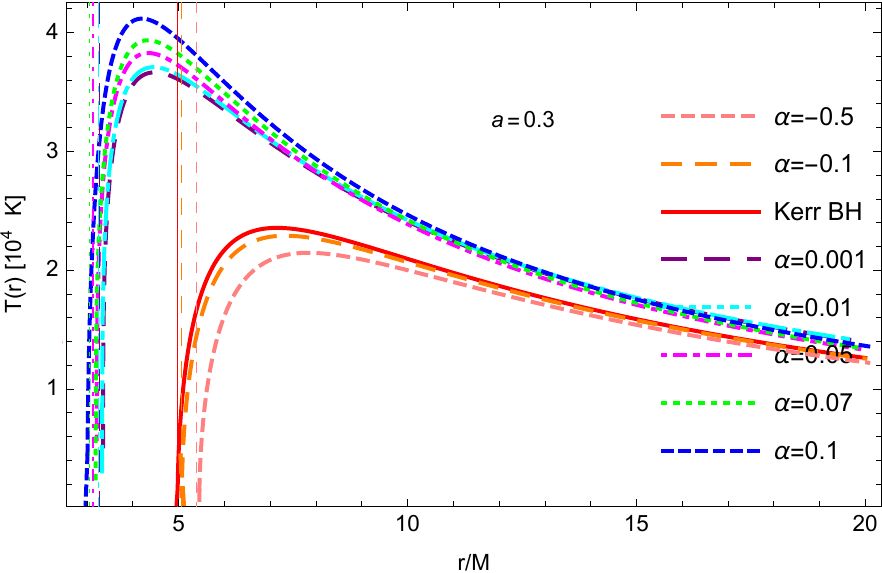}
\caption{\footnotesize The disk temperature $T(r)$ for a rotating BH in EGB gravity with mass accretion rate $\dot{M}=2\times10^{-6}M_{\odot}\rm yr^{-1}$, for different values of $\alpha$. The rotation parameter is set to $a=0.1$ (left panel) and $a=0.3$ (right panel), respectively. In each panel the solid curve corresponds to a slowly rotating Kerr BH. The vertical lines represent the location of the ISCO in each case.}
\label{Temperature}
\end{figure}

\begin{table}[H]
\centering
\caption{\footnotesize  $r_{\rm isco}$ of the accretion disk and the efficiency for rotating BHs in $4D$ EGB gravity.}
\begin{tabular}{l l l l l l}
\hline\hline
$a$&$\alpha$&$r_{+}/M$& $r_{\rm isco}/M$&$\epsilon$\\ [0.5ex]
\hline
\\
{0.1}
&--& 1.9949&5.6693&0.0606\\
\\
&0.01& 1.9899&5.2946&0.0650\\
\\
&0.05&1.9694 &5.2617&0.0653\\
\\
&0.07& 1.9589&5.2450&0.0655\\
\\
&0.1&1.9431 &5.2196&0.0657\\
 \\
{0.2}
&--&1.9798 &5.3294&0.0646\\
\\
&0.01& 1.9746&4.4583&0.0775\\
\\
&0.05&1.9534 &4.4084&0.0782\\
\\
&0.07&1.9426 &4.3826&0.0785\\
\\
&0.1& 1.9262&4.3431&0.0791\\
\\
{0.3}
&--& 1.9539&4.9786&0.0694\\
\\
&0.01&1.9484&3.2559&0.1062\\
\\
&0.05&1.9261 &3.1323&0.1095\\
\\
&0.07&1.9146& 3.0624&0.1115\\
\\
&0.1&1.8972& 2.9434&0.1149\\
\\
\hline\hline
\end{tabular}
\end{table}

In figure 6, for a fixed value of $\alpha$, we have presented the effect of the rotation parameter on the energy flux and disk temperature. As was mentioned before, in a similar fashion to Kerr BH, increasing $a$ for a  co-rotating (counter-rotating) disk causes the efficiency of a rotating EGB BH to increase (decreases), so that for $\alpha=0.07$ the efficiency increases from $6.5$\% for a=0.1 to $11\%$ for a=0.3.
We have also displayed in figure 7 the emitted flux as a function of the energy for rotating EGB BHs for different values of the BH spin. We see that the maximum flux of the disk increases as the value of the spin parameter increases and the corresponding energy is shifted to lower energies.

\begin{figure}
\centering
\includegraphics[width=3.0in]{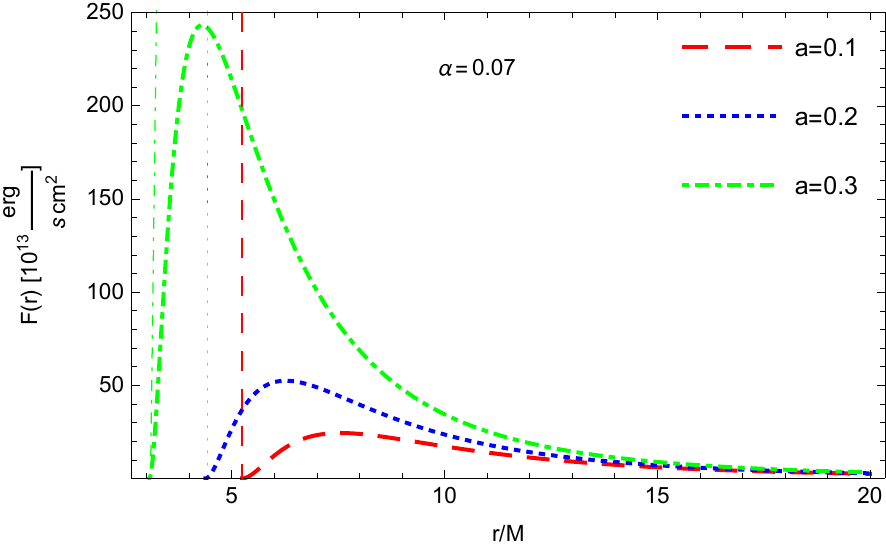}
\includegraphics[width=3.0in]{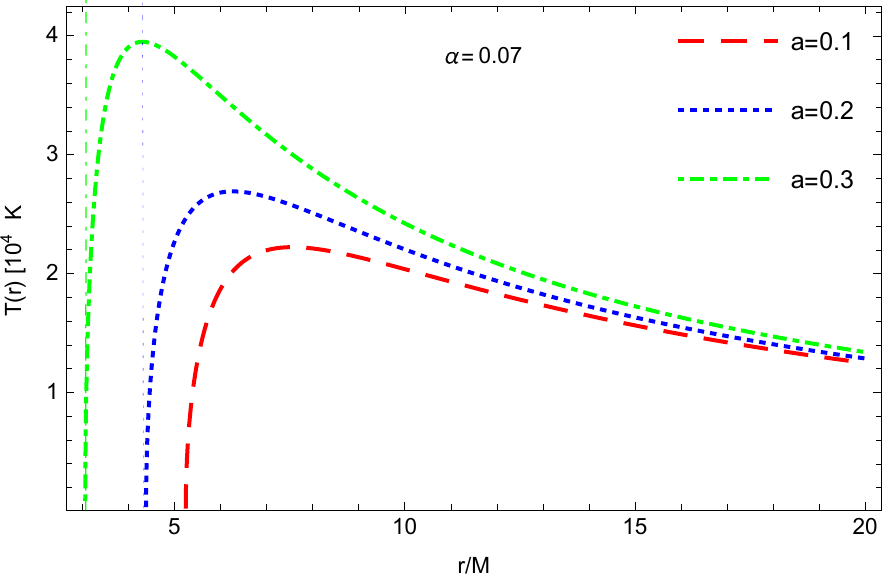}\\
\caption{\footnotesize The energy flux $F(r)$ (left panel) and  disk temperature $T(r)$ (right panel) for a rotating BH in $4D$ EGB gravity with  mass accretion rate $\dot{M}=2\times10^{-6}M_{\odot}\rm yr^{-1}$, for different values of rotation parameter $a$. The GB coupling parameter is set to $\alpha=0.07$. The vertical lines represent the location of the ISCO in each case.}
\label{temp}
\end{figure}

\begin{figure}[H]
\centering
\includegraphics[width=3.0in]{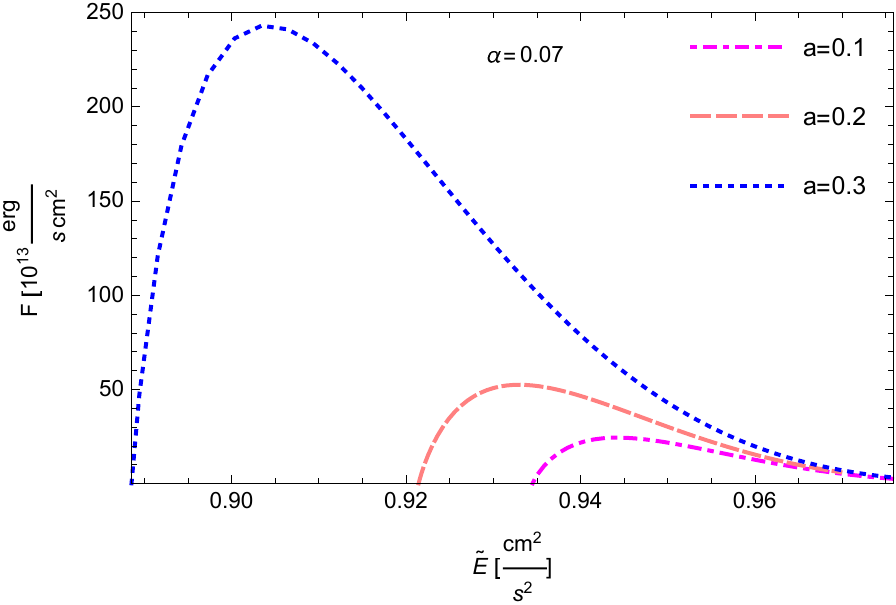}
\caption{\footnotesize The emitted flux as a function of the energy for a rotating BH in $4D$ EGB gravity. The GB coupling parameter is set to $\alpha=0.07$.}
\label{FE}
\end{figure}

\newpage
Finally, to compare our results with an actual BH-accretion disk system such as Cygnus X-1, we consider a stellar sized BH with mass $M=14.8M_{\odot}$ and an accretion rate $\dot{M}\sim10^{18}\rm g$ $ \rm s^{-1}$. It is known that the X-ray binary Cygnus X-1 contains a near-extreme Kerr BH with $0.93\leq a\leq 0.96$ \cite{cygnus1}--\cite{cygnus2}. So, in what follows we set the spin parameter to $a=0.93$ and $a=0.95$. We shall also consider the case where $a=0.75$ for a maximally rotating polytropic star \cite{pstar}.

The maximum values of the energy flux $ F_{\rm max}(r)$, disk temperature $T_{\rm max}(r)$ and $\nu L(\nu)_{\rm max}$ for rotating $4D$ EGB BHs are presented in Table 2 and compared to that of the Kerr BH. The cut-off frequency, $\nu_{\rm crit}$, for which the maximum luminosity is obtained is also given. We see that for a given value of the spin parameter, with increasing the GB coupling these maximum values also increase and the critical frequencies shift to higher values. Moreover, it is clear that by increasing the rotation parameter $a$, the differences in the maxima of the flux of rotating EGB BHs and Kerr BHs is also increasing, while for $\alpha=10^{-5}$ the rotating $4D$ EGB BHs are indistinguishable from Kerr BHs.

In figure 8, the effect of the GB coupling on the disk spectra for rotating EGB BHs is shown. Similar to the case of the energy flux and disk temperature, we see that for positive $\alpha$, the disks around rotating BHs in $4D$ EGB gravity are more luminous than the Kerr BH in GR, while for negative ones they are less luminous.

\begin{figure}[H]
\centering
\includegraphics[width=3.0in]{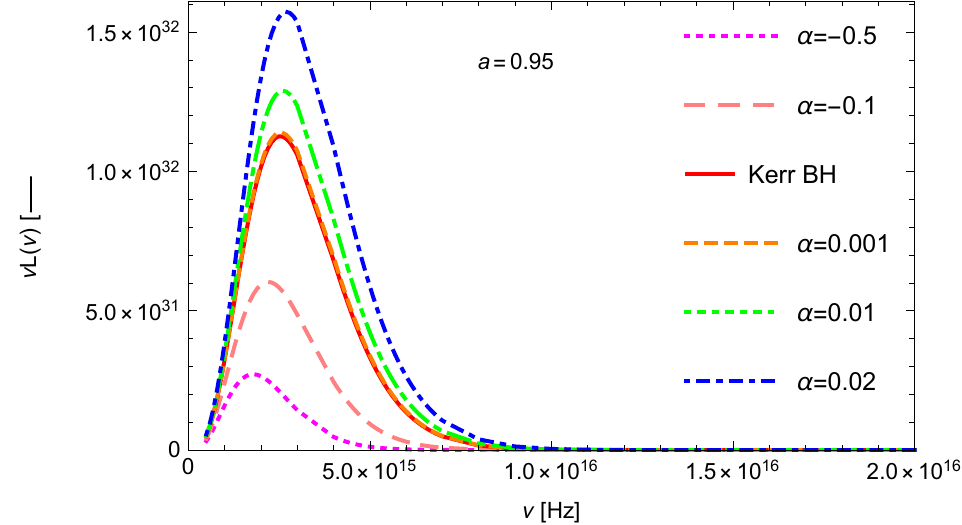}
\caption{\footnotesize The emission spectrum $\nu L(\nu)$ of the accretion disk around a rotating $4D$ EGB BH with mass accretion rate $\dot{M}\sim10^{18}\rm g$ $ \rm s^{-1}$ for different values of $\alpha$, as a function of frequency $\nu$. The solid curve represents the disk spectrum for a Kerr BH.}
\label{Luminosity}
\end{figure}

\begin{table}[H]
\centering
\caption{ The maximum values of the radiant energy flux $F(r)$, temperature distribution $T(r)$ and the emission spectra. The cut-off frequency is also shown in last column.}
\begin{tabular}{l l l l l l l}
\hline\hline
$a$&$\alpha$&  $F_{\rm max}$ [\rm erg $\rm s^{-1}$ $\rm cm^{-2}$]$\times 10^{13}$&   $T_{\rm max}$ [\rm K] $\times 10^{4}$   &$\nu L(\nu)_{\rm max}$ [\rm erg $\rm s^{-1}$]$\times 10^{31}$    &$\nu_{\rm crit}$[\rm Hz]$\times 10^{15}$\\ [0.5ex]
\hline
{0.75}
&-- & 5.1338 & 3.0847 & 3.0247 & 1.8302\\
\\
&0.0001 &5.1355&3.0849& 3.0261 &1.8389\\
\\
&0.001&5.1510& 3.0873&3.0340&1.8391\\
\\
&0.01&5.3121&3.1112&3.0829&1.8418\\
\\
{0.93}
&--& 3.4122$\times 10$ &4.9529&9.0928&2.4125\\
\\
&0.0001&3.4180$\times 10$ &4.9551&9.1276&2.4172\\
\\
&0.001&3.4716$\times 10$ &4.9743&9.1933&2.4312\\
\\
&0.01&4.1273$\times 10$ &5.1942&9.9877&2.5053\\
\\
{0.95}
&--&5.1280$\times 10$ &5.4839& 1.1267$\times 10$ & 2.5307\\
\\
&0.0001&5.1409$\times 10$ &5.4874&1.1281$\times 10$ &2.5610\\
\\
&0.001&5.2589$\times 10$ &5.5186&1.1385$\times 10$ &2.5630\\
\\
&0.01&6.8805$\times 10$ &5.9021&1.2898$\times 10$  &2.6289\\
\hline\hline
\end{tabular}
\end{table}

\section{Conclusions}
In this paper we have studied  electromagnetic properties of thin accretion disks around rotating $4D$ EGB BHs using Novikov-Thorne model. We have numerically solved the ISCO equation and found that by increasing the GB coupling the ISCO radius for rotating EGB BHs decreases. We then investigated the effect of the GB coupling parameter $\alpha$ for a fixed value of the rotation parameter $a$, on the energy flux, temperature distribution, luminosity spectra and energy conversion efficiency of thin disks and showed that with increasing  $\alpha$, all quantities also increase. We have also compared thin accretion disk properties with rotating Kerr BH in GR. We found that for the same value of the rotation parameter and for positive $\alpha$, the ISCO radius of rotating $4D$ EGB BHs is smaller than that of the Kerr BH. This result is to be expected because the GB term, as a candidate for dark energy, weakens the strength of gravity and thus the ISCO radius takes smaller values. However, for negative values of the GB coupling constant $\alpha$ tha ISCO radius of rotating $4D$ EGB BHs is larger than that of the Kerr BH. Also, similar to the slowly rotating Kerr BH, by increasing the rotational velocity, the energy flux, temperature distribution, luminosity spectra and energy conversion efficiency increase for a fixed value of $\alpha$. However, the rate of this increase becomes larger for EGB rotating BHs. Finally, by considering a stellar sized mass BH and calculating the observable characteristics of thin disks including the maximum values of $F_{\rm max}(r)$, $T_{\rm max}(r)$ and $\nu L(\nu)_{\rm max}$, we showed that thin accretion disks around rotating $4D$ EGB BHs are hotter and more luminous than in GR for positive $\alpha$, while they are cooler and less luminous for negative $\alpha$. It was also found that for $\alpha=10^{-5}$ the rotating EGB BHs are indistinguishable from Kerr BHs.

\section*{Acknowledgements}
We would like to thank the anonymous referee for valuable comments.


\begin{thebibliography}{99}

\bibitem{Shakura} N. I. Shakura and R. A. Sunyaev, {\it Astron. Astrophys} {\bf 24} (1973) 33.
\bibitem{Novikov} I. D. Novikov and K. S. Thorne, Astrophysics and black holes, in {\it Black Holes}, edited by C. De Witt and B. De Witt (Gordon and Breach, New York, 1973).
\bibitem{Page} D. N. Page and K. S. Thorne, {\it Astrophys. J} {\bf 191} (1974) 499.
\bibitem{Thorne} K. S. Thorne, {\it Astrophys. J} {\bf 191} (1974) 507.

\bibitem{FR1} C. S. J. Pun, Z. Kov\'{a}cs, T. Harko, {\it Phys. Rev.} D {\bf 78} (2008) 024043.
\bibitem{FR2} D. Perez, G. E. Romero and S. E. Perez Bergliaffa, {\it  Astron. Astrophys} {\bf 551} (2013) A4.
\bibitem{FR3} K. V. Staykov, D. D. Doneva and S. S. Yazadjiev, {\it JCAP} {\bf 2016} (2016) 061.

\bibitem{SVT} D. Perez, F. G. L. Armengol and G. E. Romero, {\it Phys. Rev.} D {\bf 95} (2017) 104047.

\bibitem{EMd1} R. Kh. Karimov, R. N. Izmailov, A. Bhattacharya and K. K. Nandi, {\it Eur. Phys. J.} C {\bf 78} (2018) 788.
\bibitem{EMd2} M. Heydari-Fard, M. Heydari-Fard and H. R. Sepangi, {\it Eur. Phys. J.} C {\bf 80} (2020) 351.

\bibitem{EdGB1} H. Zhang, M. Zhou, C. Bambi, B. Kleihaus, J. Kunz and E. Radu, {\it Phys. Rev.} D {\bf 95} (2017) 104043.
\bibitem{EdGB2} M. Heydari-Fard and H. R. Sepangi, {\it Phys. Lett.} B {\bf 816} (2021) 136276.


\bibitem{Chern} T. Harko, Z. Kov\'{a}cs and F. S. N. Lobo, {\it Class. Quant. Grav} {\bf 27} (2010) 105010.

\bibitem{Horava} T. Harko, Z. Kov\'{a}cs and F. S. N. Lobo, {\it Phys. Rev.} D {\bf 80} (2009) 044021.

\bibitem{Kaluza} S. Chen and J. Jing, {\it Phys. Lett.} B {\bf 704} (2011) 641.
\bibitem{brane1} C. S. J. Pun, Z. Kovacs and T. Harko, {\it Phys. Rev.} D {\bf 78} (2008) 084015.
\bibitem{brane2} M. Heydari-Fard, {\it Class. Quant. Grav.} {\bf 27} (2010) 235004.


\bibitem{WH1} T. Harko, Z. Kov\'{a}cs and F. S. N. Lobo, {\it Phys. Rev.} D {\bf 79} (2009) 064001.
\bibitem{WH2} R. Kh. Karimov, R. N. Izmailov and K. K. Nandi, {\it Eur. Phys. J.} C {\bf 79} (2019) 952.
\bibitem{s1} Z. Kov\'{a}cs, K. S. Cheng and T. Harko, {\it Astron. Astrophys} {\bf 500} (2009) 621.
\bibitem{s2} F. S. Guzman, {\it Phys. Rev.} D {\bf 73} (2006) 021501.
\bibitem{s3} Y. F. Yuan, R. Narayan and M. J. Rees, {\it Astrophys. J} {\bf 606} (2004) 1112.
\bibitem{nk1} Z. Kov\'{a}cs and T. Harko, {\it Phys. Rev.} D {\bf 82} (2010) 124047.
\bibitem{nk2} P. S. Joshi, D. Malafarina and R. Narayan, {\it Class. Quant. Grav} {\bf 31} (2014) 015002.
\bibitem{nk3} S. Shahidi, T. Harko and Z. Kov\'{a}cs, {\it Eur. Phys. J.} C {\bf 80} (2020) 162.

\bibitem{EGB} C. Liu, T. Zhu and Q. Wu, {\it Chin. Phys.} C {\bf 45} (2021) 015105.
%&&&&&&&&&&&&&&&&&&&&&&&&&&&&&&&&&&&&&&&&&&&&&&&&&&&
\bibitem{Kanti} P. Kanti, N. E. Mavromatos, J. Rizos, K. Tamvakis and E. Winstanley, {\it Phys. Rev.} D {\bf 54} (1996) 5049.

\bibitem{Glavan} D. Glavan and C. Lin, {\it Phys. Rev. Lett} {\bf 124} (2020) 081301.

\bibitem{Cai} R. G. Cai, L. M. Cao and N. Ohta, {\it JHEP} {\bf 1004} (2010) 082.

\bibitem{c1} W. Y. Ai, {\it Commun. Theor. Phys} {\bf 72} (2020) 095402.
\bibitem{c2} M. Gurses, T. C. Sisman and B. Tekin, {\it Eur. Phys. J.} C {\bf 80} (2020) 647.
\bibitem{c3} S. Mahapatra, {\it Eur. Phys. J.} C {\bf 80} (2020) 992.
\bibitem{c4} F. W. Shu, {\it Phys. Lett.} B {\bf 811} (2020) 135907.
\bibitem{c5} S. X. Tian and Z. H. Zhu, arXiv: 2004.09954 [gr-qc].
\bibitem{c6} J. Bonifacio, K. Hinterbichler and L. A. Johnson, {\it Phys. Rev.} D {\bf 102} (2020) 024029.
\bibitem{c7} J. Arrechea, A. Delhom and A. Jim\'{e}nez-Cano, {\it Chin. Phys.} C {\bf 45} (2021) 013107.

\bibitem{n1} H. L\"{u} and Y. Pang, {\it Phys. Lett.} B {\bf 809} (2020) 135717.
\bibitem{n2} T. Kobayashi, {\it JCAP} {\bf 07} (2020) 013.
\bibitem{n3} P. G. S. Fernandes, P. Carrilho, T. Clifton and D. J. Mulryne, {\it Phys. Rev.} D {\bf 102} (2020) 024025.
\bibitem{n4} R. A. Hennigar, D. Kubiz\v{n}\'{a}k, R. B. Mann and C. Pollack, {\it JHEP} {\bf 2020} (2020) 27.
\bibitem{n5} K. Aoki, M. A. Gorji and S. Mukohyama, {\it Phys. Lett.} B {\bf 810} (2020) 135843.

\bibitem{charge} P. G. S. Fernandes, {\it Phys. Lett.} B {\bf 805} (2020) 135468.
\bibitem{rotating1} R. Kumar and S. G. Ghosh, {\it JCAP} {\bf 2020} (2020) 053.
\bibitem{rotating2} S. W. Wei and Y. X. Liu, {\it Eur. Phys. J. Plus} {\bf 136} (2021) 436.
\bibitem{love1} R. A. Konoplya and A. Zhidenko, {\it Phys. Rev.} D {\bf 101} (2020) 084038.
\bibitem{love2} A. Casalino, A. Coll\'{e}aux, M. Rinaldi and S. Vicentini, {\it Phys. Dark. Univ} {\bf 31} (2021) 100770.
\bibitem{string} D. V. Singh, S. G. Ghosh and S. D. Maharaj, {\it Phys. Dark. Univ} {\bf 30} (2020) 100660.
\bibitem{bardeen} A. Kumar and R. Kumar, arXiv:2003.13104 [gr-qc].
\bibitem{hayward} A. Kumar and S. G. Ghosh, arXiv:2004.01131 [gr-qc].
\bibitem{wormhole1} K. Jusufi, A. Banerjee and S. G. Ghosh, {\it Eur. Phys. J.} C {\bf 80} (2020) 698.
\bibitem{wormhole2} P. Liu, C. Niu, X. Wang and C. Y. Zhang, arXiv:2004.14267 [gr-qc].
\bibitem{star} D. D. Doneva and S. S. Yazadjiev, arXiv:2003.10284 [gr-qc].

\bibitem{isco} M. Guo and P. C. Li, {\it Eur. Phys. J.} C {\bf 80} (2020) 588.
\bibitem{test} Y. P. Zhang, S. W. Wei and Y. X. Liu, {\it Universe} {\bf 6} (2020) 103.
\bibitem{shadow1} R. Roy and S. Chakrabarti, {\it Phys. Rev.} D {\bf 102} (2020) 024059.
\bibitem{shadow2} X. X. Zeng, H. Q. Zhang and H. Zhang, {\it Eur. Phys. J.} C {\bf 80} (2020) 872.

\bibitem{lensing1} S. U. Islam, R. Kumar and S. G. Ghosh, {\it JCAP} {\bf 09} (2020) 030.
\bibitem{lensing2} R. Kumar, S. U. Islam and S. G. Ghosh, {\it Eur. Phys. J.} C {\bf 80} (2020) 1128.
\bibitem{lensing3} M. Heydari-Fard, M. Heydari-Fard and H. R. Sepangi, arXiv:2004.02140 [gr-qc].
\bibitem{lensing4} X. H. Jin, Y. X. Gao and D. J. Liu, {\it Int. J. Mod. Phys.} D {\bf 29} (2020) 2050065.

\bibitem{QNM1} R. A. Konoplya and A. F. Zinhailo, {\it Eur. Phys. J.} C {\bf 80} (2020) 1049.
\bibitem{QNM2} M. S. Churilova, {\it Phys. Dark. Univ} {\bf 31} (2021) 100748.
\bibitem{QNM3} A. K. Mishra, {\it Gen. Relat. Grav} {\bf 52} (2020) 106.
\bibitem{QNM4} A. Arag\'{o}n, R. B\'{e}car, P. A. Gonz\'{a}lez and Y. V\'{a}squez, {\it Eur. Phys. J.} C {\bf 80} (2020) 773.
\bibitem{QNM5} S. Devi, R. Roy and S. Chakrabarti, {\it Eur. Phys. J.} C {\bf 80} (2020) 760.

\bibitem{stability1} R. A. Konoplya and A. Zhidenko, {\it Phys. Dark. Univ} {\bf 30} (2020) 100697.
\bibitem{stability2} M. A. Cuyubamba, {\it Phys. Dark. Univ} {\bf 31} (2021) 100789.
\bibitem{stability3} P. Liu, C. Niu and C. Y. Zhang, {\it Chin. Phys.} C {\bf 45} (2021) 2.

\bibitem{th1} S. A. Hosseini Mansoori, {\it Phys. Dark. Univ} {\bf 31} (2021) 100776.
\bibitem{th2}  K. Hegde, A. N. Kumara, C. L. A. Rizwan, K. M. Ajith and M. S. Ali, arXiv:2003.08778 [gr-qc].
\bibitem{th3} S. Ying, {\it Chin. Phys.} C {\bf 44} (2020) 125101.
\bibitem{th4} D. V. Singh and S. Siwach, {\it Phys. Lett.} B {\bf 808} (2020) 135658.

\bibitem{Hawking1} C. Y. Zhang, P. C. Li and M. Guo, {\it Eur. Phys. J.} C {\bf 80} (2020) 874.
\bibitem{Hawking2} R. A. Konoplya and A. F. Zinhailo, {\it Phys. Lett.} B {\bf 810} (2020) 135793.

\bibitem{star1} A. Banerjee and K. N. Singh, {\it Phys. Dark. Univ} {\bf 31} (2021) 100792.
\bibitem{star2} A. Banerjee, T. Tangphati and P. Channuie, {\it Astrophys. J} {\bf 909} (2021) 14.


\bibitem{a1} R. A. Konoplya and A. Zhidenko, {\it Phys. Rev} D {\bf 102} (2020) 064004.
\bibitem{a2} K. Yang, B. M. Gu, S. W. Wei and Y. X. Liu, {\it Eur. Phys. J.} C {\bf 80} (2020) 662.

\bibitem{a3} A. Abdujabbarov, J. Rayimbaev, B. Turimov and F. Atamurotov, {\it Phys. Dark. Univ} {\bf 30} (2020) 100715.
\bibitem{a4} X. H. Ge and S. J. Sin, {\it Eur. Phys. J.} C {\bf 80} (2020) 695.
\bibitem{a5} C. Y. Zhang, S. J. Zhang, P. C. Li and M. Guo, {\it JHEP} {\bf 08} (2020) 105.
\bibitem{a6} B. Eslam Panah, Kh. Jafarzade and S. H. Hendi, {\it Nucl. Phys.} B {\bf 961} (2020) 115269.

\bibitem{a7} Z. Haghani, {\it Phys. Dark. Univ} {\bf 30} (2020) 100720.
\bibitem{a8} H. Mohseni Sadjadi, {\it Phys. Dark. Univ} {\bf 30} (2020) 100728.
\bibitem{a9} S. L. Li, P. Wu and H. Yu, arXiv:2004.02080 [gr-qc].
\bibitem{a10} G. Narain and H. Q. Zhang, arXiv:2005.05183 [gr-qc].

\bibitem{a11} D. Malafarina, B. Toshmatov and N. Dadhich, {\it Phys. Dark. Univ} {\bf 30} (2020) 100598.
\bibitem{a12} N. Dadhich, {\it Eur. Phys. J.} C {\bf 80} (2020) 832.

\bibitem{a13} T. Clifton, P. Carrilho, P. G. S. Fernandes and D. J. Mulryne, {\it Phys. Rev.} D {\bf 102} (2020) 084005.
\bibitem{a14} J. X. Feng, B. M. Gu and F. W. Shu, {\it Phys. Rev.} D {\bf 103} (2021) 064002.
\bibitem{a15} M. A. Garc\'{\i}a-Aspeitia and A. Hern\'{a}ndez-Almada, {\it Phys. Dark. Univ} {\bf 32} (2021) 100799.


\bibitem{Torres} D. Torres, {\it Nucl. Phys.} B {\bf 626} (2002) 377.


\bibitem{Bardeen} J. M. Bardeen, W. H. Press and A. S. Teukolsky, {\it Astrophys. J} {\bf 178} (1972) 347.


\bibitem{book} C. Bambi, Black Holes: A Laboratory for Testing Strong Gravity, Springer, Singapore (2017)

\bibitem{ed1} J. E. McClintock, R. Shafee, R. Narayan, R. A. Remillard, S. W. Davis and L. X. Li, {\it ArstroPhys. J} {\bf 652} (2006) 518.
\bibitem{ed2} R. F. Penna, J. C. McKinney, R. Narayan, A. Tchekhovskoy, R. Shafee and J. E. McClintock, {\it Mon. Not. Roy. Astron. Soc} {\bf 408} (2010) 752.
\bibitem{ed3} J. F. Steiner, J. E. McClintock, R. A. Remillard, L. Gou, S. Yamada and R. Narayan, {\it ArstroPhys. J} {\bf 718} (2010) L117.
\bibitem{ed4} A. K. Kulkarni et al., {\it Mon. Not. Roy. Astron. Soc} {\bf 414} (2011) 1183.

\bibitem{sgr} Q. D. Wang et al., {\it Science } {\bf 341} (2013) 981.


\bibitem{cygnus1} L. Gou et al., {\it Astrophys. J.} {\bf 742} (2011) 85.
\bibitem{cygnus2} D. J. Walton et al., {\it Astrophys. J.} {\bf 826} (2016) 87.
\bibitem{pstar} S. L. Shapiro and M. Shibata, {\it Astrophys. J.} {\bf 577} (2002) 904.

\end{thebibliography}
\end{document}